%% file: main.tex
\documentclass[sigconf]{acmart}

\renewcommand\footnotetextcopyrightpermission[1]{} % removes footnote with conference information in first column

\usepackage{amsmath,amssymb,amsfonts}
\usepackage{graphicx}
\usepackage{textcomp}
\usepackage{xcolor}
\usepackage{makecell}

\usepackage{multirow}
\usepackage{siunitx}
\usepackage{geometry}
\usepackage{booktabs,siunitx,adjustbox}
\usepackage{diagbox}

\usepackage{multirow}
\usepackage{tabularx}
\usepackage{algorithm,algorithmicx,algpseudocode}
\usepackage{hyperref}
\usepackage{comment}
\usepackage{enumitem}
\usepackage{array}
\usepackage{subcaption}
\usepackage{pifont}
\usepackage{microtype}
\usepackage{booktabs}    % 三线表
\usepackage{caption}
\usepackage{setspace} % 会影响表格中的字体大小
\AtBeginDocument{%
  }

\acmConference[Accepted to ICCAD 2026]{IEEE/ACM International Conference on Computer-Aided Design}{November 08--12, 2026}{San Jose, CA, USA}

\begin{document}

\title[PICopilot]{PICopilot: An LLM-based Agentic Framework for Assisting Photonic Integrated Circuit Design via Script Generation}

\author{\parbox{\linewidth}{\centering
  \fontsize{12}{15}\selectfont Xiaohan Jiang$^{1}$, Zeyu Li$^{1}$, Wei Zhang$^1$, Jiang Xu$^{2,*}$ \\ \vspace{2pt}
  \fontsize{9}{11}\selectfont $^1$Department of Electronic and Computer Engineering, The Hong Kong University of Science and Technology \\ 
  \fontsize{9}{11}\selectfont $^2$Microelectronics Thrust, The Hong Kong University of Science and Technology (Guangzhou) \\
  \fontsize{9}{11}\selectfont $^*$Corresponding author: jiang.xu@hkust-gz.edu.cn
}}

\begin{abstract}
    % v7
    The rapid development of photonic integrated circuits (PICs) is shifting the design flow from traditional graphical user interface (GUI)-based methods to script-based methods for higher flexibility, portability, and maintainability. However, script-based design introduces new challenges, requiring designers to possess additional proficiency in tool application programming interfaces (APIs) and programming. It also demands greater effort and time because it is inherently less intuitive and more complex than GUI-based methods. As PICs grow in scale and complexity, the productivity gap between design needs and manual scripting capabilities continues to widen.
    To address this gap, we introduce PICopilot, the first large language model (LLM)-based agentic framework that assists in PIC design via automated design script generation from natural language instructions. PICopilot leverages a multi-agent architecture with a feedback mechanism and a specifically designed retrieval-augmented generation (RAG) pipeline, achieving a high success rate and reliability. Experimental results on a benchmark of diverse PIC scripting tasks demonstrate that PICopilot successfully completes all 48 tasks and outperforms other LLM-based approaches without incurring substantial extra latency or cost, even solving 21 more tasks than the advanced GPT-5 model with a general RAG pipeline.
\end{abstract}

\begin{CCSXML}
<ccs2012>
<concept>
<concept_id>10010583.10010786.10010787.10010791</concept_id>
<concept_desc>Hardware~Emerging tools and methodologies</concept_desc>
<concept_significance>500</concept_significance>
</concept>
<concept>
<concept_id>10010583.10010682.10010712.10010715</concept_id>
<concept_desc>Hardware~Software tools for EDA</concept_desc>
<concept_significance>500</concept_significance>
</concept>
<concept>
<concept_id>10010583.10010786.10010810</concept_id>
<concept_desc>Hardware~Emerging optical and photonic technologies</concept_desc>
<concept_significance>300</concept_significance>
</concept>
</ccs2012>
\end{CCSXML}

\ccsdesc[500]{Hardware~Emerging tools and methodologies}
\ccsdesc[500]{Hardware~Software tools for EDA}
\ccsdesc[300]{Hardware~Emerging optical and photonic technologies}

\keywords{Photonic Design Automation, Photonic Integrated Circuits, Large Language Models, Retrieval-Augmented Generation, Agents}

% control the author name in the header
\renewcommand{\shortauthors}{Xiaohan Jiang, Zeyu Li, Wei Zhang and Jiang Xu}

% control the author name in the ACM Reference Format
\makeatletter
\xdef\authors{Xiaohan Jiang, Zeyu Li, Wei Zhang, and Jiang Xu}
\makeatother

\maketitle

\input{Introduction}
\input{Preliminary}
\input{Framework}

\input{Experiment}

\input{Conclusion}

\clearpage
\bibliographystyle{ACM-Reference-Format}
\bibliography{ref}
\end{document}

%% file: Introduction.tex
\section{Introduction}
Photonic integrated circuits (PICs) are rapidly emerging as a key technology for next-generation computing and communication systems, offering superior power efficiency, bandwidth, and speedup \cite{ning2024photonic}. Advances in manufacturing processes have greatly increased their integration density and scale \cite{siew2021review}, enabling the design of large and complex PICs \cite{ashtiani2022chip,xu2024large,bandyopadhyay2024single,ahmed2025universal}.

\begin{figure}[t]
\centerline{\includegraphics[width=\linewidth]{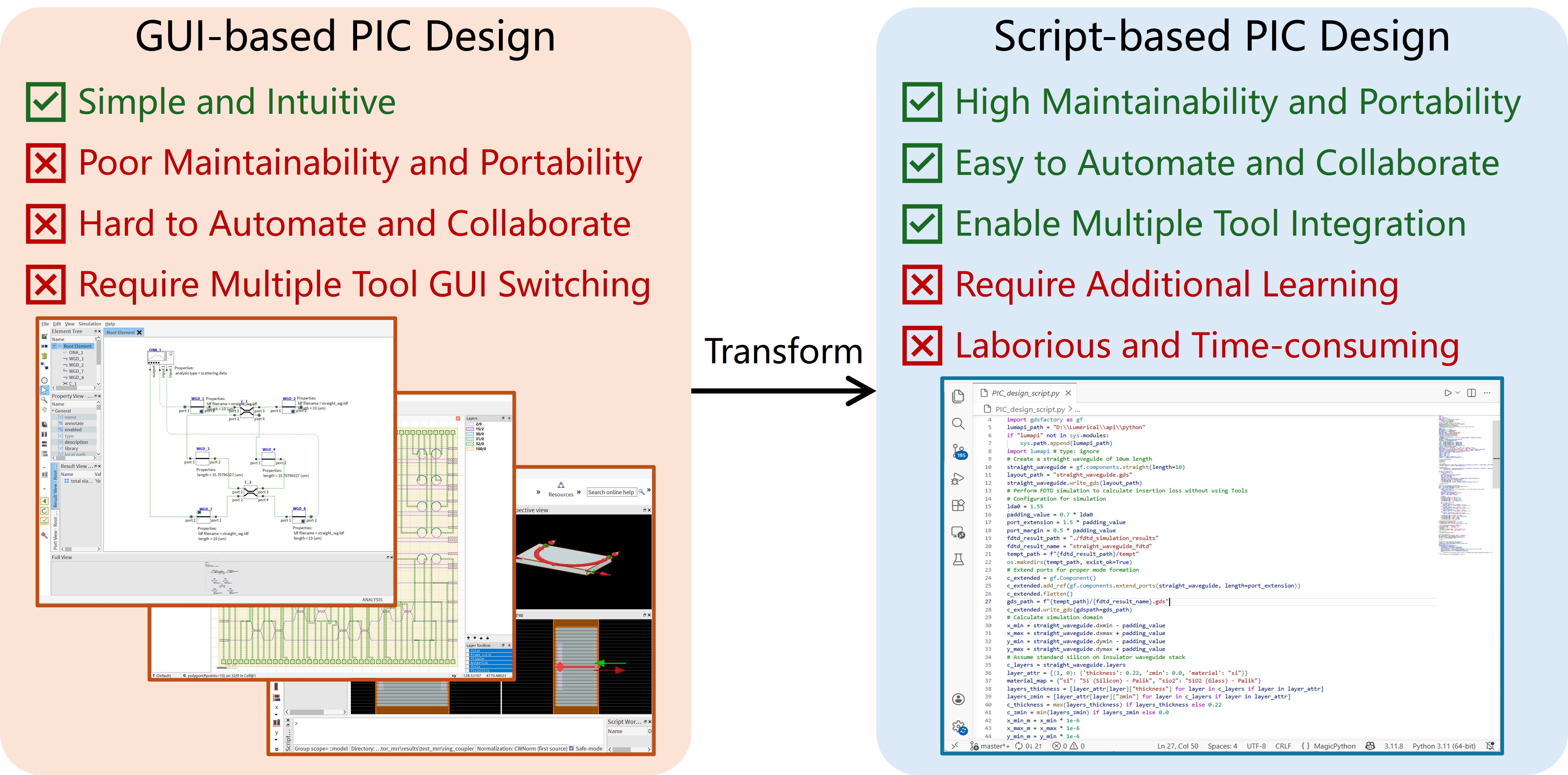}}
\vspace*{-0.45\baselineskip}
\caption{An illustration of the GUI-based PIC design and the script-based PIC design.}
\label{fig:GUI_vs_Script}
\vspace*{-0.1\baselineskip}
\end{figure}

In PIC design, since a mature end-to-end design tool is still lacking, designers have to use various function-specific tools to complete the entire design flow, including layout design tools \cite{Gdsfactory, Luceda, Leditor, kofferlein2020klayout}, verification tools \cite{kofferlein2020klayout, checkmatedrc}, specific simulators for different evaluation metrics \cite{INTERCONNECT, FDTD, MODE, tidy3D, OptSim}, and emerging design automation tools \cite{Liu2013fiona, jiang2025picelf, chen2025bi, wu2025constraints}.
Traditionally, designers utilize these tools through their graphical user interfaces (GUIs), which have long been the only operating mode they supported. As shown in Figure~\ref{fig:GUI_vs_Script}, this GUI‑based design flow offers a simple and intuitive user experience. However, it poses significant challenges in portability, maintainability, and collaborative development. Design configurations of the flow are typically scattered across different interfaces, rendering them opaque and hindering readability, version control, and sharing. Furthermore, such a flow is also difficult to automate, particularly when dealing with repetitive manual operations. To overcome these limitations, the PIC community is increasingly adopting script-based design methodologies as PIC design tools evolve, where designers execute and control design flows by writing code scripts. As illustrated in Figure~\ref{fig:GUI_vs_Script}, this design paradigm facilitates the capture of design intent, enables seamless integration between tools, and enhances flexibility, reproducibility, and maintainability.

% \begin{figure}[htbp]
% % \vspace*{-0.3\baselineskip}
% \vspace*{-0.3\baselineskip}
% \centerline{\includegraphics[width=\linewidth]{figure/GUI_vs_Script_latest.jpg}}
% % \vspace{-4pt}
% \vspace*{-0.3\baselineskip}
% \caption{An illustration of the GUI-based PIC design and the script-based PIC design.}
% \label{fig:GUI_vs_Script}
% % \vspace{-16pt}
% % \vspace*{-0.45\baselineskip}
% \vspace*{-0.45\baselineskip}
% \end{figure}

% \begin{figure}[htbp]
% \vspace*{-0.65\baselineskip}
%     \centering
%     \captionsetup[subfigure]{skip=3pt}
%     \begin{minipage}[c]{\linewidth}
%         \centering
%         \includegraphics[width=\linewidth]{figure/GUI_vs_Script_latest.jpg}
%         % \vspace*{-0.5\baselineskip}
%         \subcaption{GUI-based PIC design vs. Script-based PIC design}
%         \label{fig:pm_0}
%     \end{minipage} \\
%     \vspace{0.2em}
%     \begin{minipage}[c]{\linewidth}
%         \centering
%         \includegraphics[width=0.75\linewidth]{figure/manual_vs_llm_resize.jpg}
%         % \vspace*{-0.3\baselineskip}
%         \subcaption{Core innovation of this work}
%         \label{fig:pm_1}
%     \end{minipage} \\
%     \vspace*{-0.65\baselineskip}
%     \caption{An illustration of the PIC design paradigms and our core innovation.}
%     \label{fig:GUI_vs_Script}
% \vspace*{-0.65\baselineskip}
% \end{figure}

However, this script-based design flow imposes a considerable burden on PIC designers, requiring them to develop additional proficiency in both the application programming interfaces (APIs) of various design tools and general programming skills. It also lacks the intuitiveness and interactivity offered by GUI-based methods. As a result, designers often devote substantial time and effort to laborious script writing rather than focusing on the PIC design itself. As PICs continue to scale in size and complexity, scripting has become a critical time bottleneck, severely reducing current PIC design efficiency. Therefore, there is an urgent need for an automated script generation tool that can relieve PIC designers from tedious script-writing tasks, allowing them to focus on design innovation and significantly improving productivity.

Emerging large language models (LLMs) present a promising opportunity to address the aforementioned challenges. In the electronic design automation (EDA) domain, various LLM-based tools have been developed \cite{pan2025survey}. Some of these tools have achieved notable success in generating design scripts directly from natural language descriptions \cite{liu2023chipnemo,wu2024chateda, sun2025anasizecoder, wang2025mcp4eda, liu2025layoutcopilot, lai2025analogcoder, lai2025analogcoderpro}, demonstrating the feasibility and efficiency improvement of integrating LLMs into script-based design flows. Nevertheless, the application of LLMs in photonic design automation (PDA) remains limited and focuses primarily on using them to directly generate PIC designs. \cite{liu2024towards} proposed an LLM-based framework that automatically generates PIC devices based on natural language descriptions. \cite {wu2025picbench} introduced the first benchmark for LLM-automated PIC design and applied LLMs to directly design PIC by generating netlists in JSON format. \cite{sharma2025ai} developed a multi-agent framework that uses LLMs to produce domain-specific language scripts of high-level PIC designs and invokes a predefined toolchain to generate layouts. Although these studies highlight the potential of applying LLMs in PDA, they focus solely on using them to directly design PICs, rather than generating design tool scripts to assist in PIC design. Currently, there is still a significant gap in exploring how to leverage the powerful programming capabilities of LLMs to automate labor-intensive and time-consuming scripting tasks across the entire PIC design flow.

Our major contributions can be summarized as follows:
\begin{itemize}[nosep]
    \item We introduce \textbf{PICopilot}, an automated framework that utilizes LLMs' powerful coding capabilities to transform designers' natural language descriptions into executable scripts, thereby assisting PIC design. To our knowledge, it is the first tool to pioneer script generation in the PDA domain.
    \item We propose an agentic architecture with a feedback mechanism, in which multiple tailored LLM agents collaborate to complete PIC design script generation tasks. This design enhances the success rate of generating correct scripts while improving overall reliability.
    \item We develop a retrieval-augmented generation (RAG) pipeline specifically tailored for PIC design scripting tasks. It employs a high-precision retrieval paradigm that mimics the practical retrieval process of human PIC designers and utilizes a highly scalable multi-database structure. This pipeline allows existing LLMs to generate accurate scripts in the unfamiliar PIC design domain, while outperforming the general RAG pipeline used by existing related methods.
    \item We establish a comprehensive benchmark covering a wide range of real-world PIC design script writing tasks. Experimental results show that PICopilot can successfully generate functionally correct scripts for all 48 tasks, whereas other LLM-based methods, even those using the advanced GPT-5 model, can only complete a maximum of 27 tasks. Furthermore, PICopilot incurs no significant additional time or cost compared to baseline methods, rendering it highly practical.
\end{itemize}

The rest of this paper is organized as follows. Section 2 discusses the preliminaries. Section 3 details the PICopilot framework. Section 4 reports the experimental results. Section 5 gives our conclusion. 

%% file: Preliminary.tex
\section{Preliminary}

\begin{table}[t]
% \vspace*{-0.45\baselineskip}
\centering
% \caption{Comparison of LLM-based circuit design script generation tools.}
\caption{Comparison of LLM-based tools for circuit design script generation.}
% \vspace{-10pt}
\vspace*{-0.65\baselineskip}
\footnotesize
% \small
% \setlength{\tabcolsep}{5pt}
\setlength{\tabcolsep}{5pt}
\renewcommand{\arraystretch}{1.1}
\begin{tabular}{l|cc}
\toprule
\makecell[c]{\textbf{Tools}} & \textbf{Circuit} & \textbf{Method} \\
\midrule
ChipNeMo \cite{liu2023chipnemo} & Digital  & Training-based \\
ChatEDA \cite{wu2024chateda}  & Digital  & Training-based \\
DRC-Coder \cite{chang2025drc} & Digital & Training-free \\
AnaSizeCoder \cite{sun2025anasizecoder} & Analog  & Training-based \\
LayoutCopilot \cite{liu2025layoutcopilot} & Analog  & Training-free \\
AnalogCoder \cite{lai2025analogcoder}  & Analog  & Training-free \\
\textbf{PICopilot} & \textbf{Photonic} & \textbf{Training-free} \\
\bottomrule
\end{tabular}
\label{tab:existing_works}
% \vspace*{-0.45\baselineskip}
\end{table}

\subsection{Script-based PIC Design Flow} \label{sec:flow}
Modern PIC design flows are increasingly leaning towards script-driven workflows rather than GUI-based interactions. Most mainstream commercial and open-source PIC design tools, such as Ansys Lumerical suite \cite{lumerical_script}, GDSFactory \cite{Gdsfactory}, and Luceda IPKSS \cite{Luceda}, already support scripting and provide comprehensive interfaces, allowing designers to programmatically execute all necessary design tools in the PIC design flow. 
Despite the availability of other scripting languages, Python has emerged as the most suitable and widely adopted choice, as almost all PIC design tools support Python-based calls. This enables PIC designers to easily invoke various tools using a single script to complete multiple PIC design steps. In addition to tool invocation within the flow, designers also need to write additional scripts for auxiliary tasks, such as data processing, file management, and automation control, which are also well-suited for Python. Therefore, Python scripts are currently the most widely used in this domain because they can cover all necessary steps in the PIC design flow, and we take Python as the default scripting language in the remainder of this paper.

Compared with GUI-based methods, the script-driven workflow offers higher flexibility, reproducibility, and scalability. Furthermore, it enables PIC designers to automate repetitive tasks, seamlessly coordinate different design tools, and maintain version-controlled design flows.
However, this design paradigm shift introduces new challenges. Because writing scripts lacks intuitiveness and requires additional learning of tool APIs and programming skills, PIC designers often spend significant time and effort on scripting, rather than concentrating on PIC design itself. This burden is further exacerbated by the fragmented PDA ecosystem, where currently no single vendor provides a complete toolchain that meets all requirements, forcing designers to use various tools from multiple vendors with disparate APIs.
With the rapid development and increasing complexity of PICs, script writing has become a major time bottleneck in the entire PIC design flow, severely limiting productivity and urgently necessitating automation solutions.

\begin{figure*}[t]
% \vspace{-8pt}
% \centerline{\includegraphics[width=0.85\linewidth]
\centerline{\includegraphics[width=0.85\linewidth]{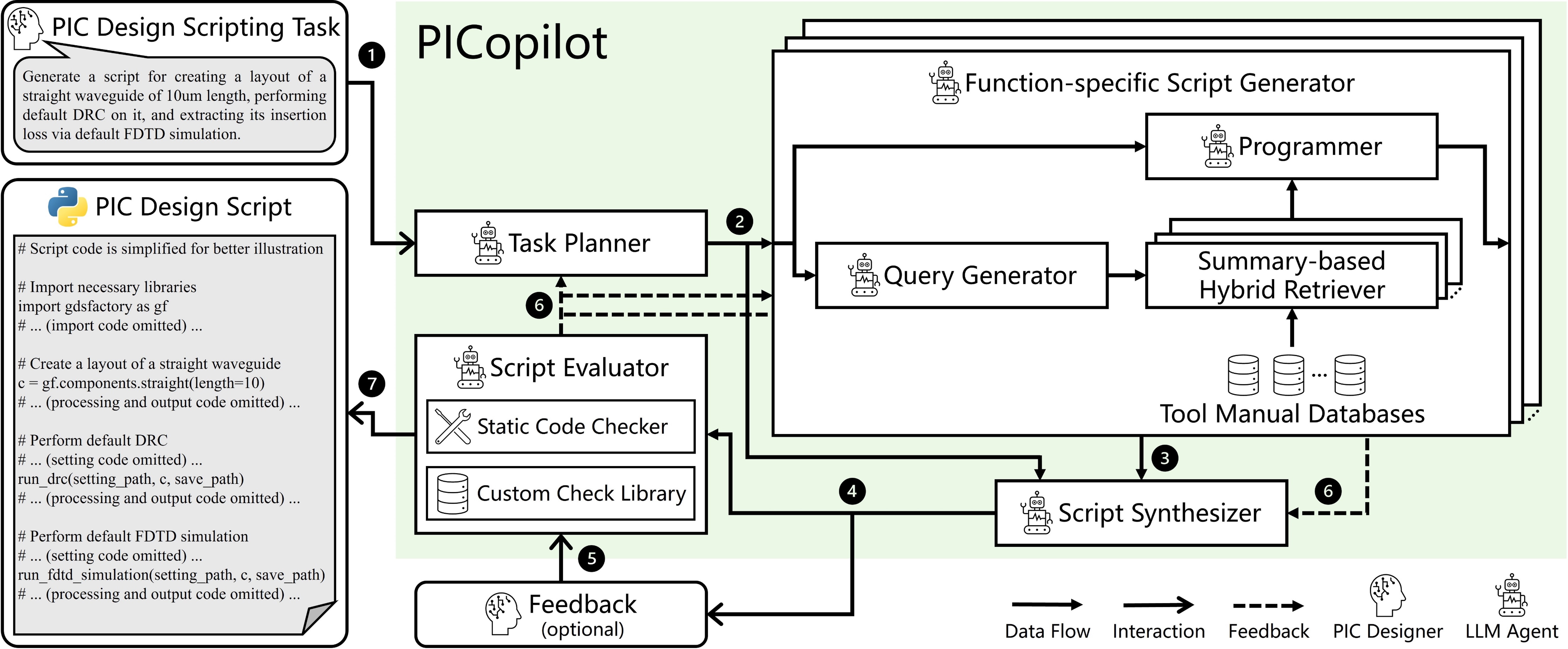}}
% \vspace{-9pt}
% \vspace*{-0.65\baselineskip}
\vspace*{-0.55\baselineskip}
\caption{Overview of PICopilot.}
\label{fig:pic_copilot}
% \vspace{-16pt}
% \vspace*{-0.65\baselineskip}
\vspace*{-0.85\baselineskip}
\end{figure*}

\subsection{LLM-based Circuit Design Script Generation}
The recent success of LLMs has brought new opportunities for automating the tedious and time-consuming scripting tasks in circuit design flows. However, current LLMs are not inherently familiar with the domain-specific scripting methods commonly used in these flows, primarily due to the scarcity of relevant data in their training corpora. This limits their ability to directly generate executable and functionally correct scripts from the designer's natural language descriptions, necessitating targeted strategies to bridge this gap and enable effective automated script generation.

Existing research in the EDA domain has investigated two main approaches to address this challenge. The first approach involves pre-training or fine-tuning LLMs on specialized datasets, helping them learn the syntax, semantics, and usage patterns of circuit design scripts \cite{liu2023chipnemo,wu2024chateda,sun2025anasizecoder}. Despite its effectiveness, this approach suffers from the scarcity of high-quality training data and incurs substantial computational and financial costs. Consequently, methods employing training-free techniques have garnered increasing attention \cite{liu2025layoutcopilot, lai2025analogcoder, chang2025drc}, among which in-context learning (ICL) \cite{dong2024survey} and retrieval-augmented generation (RAG) \cite{lewis2020retrieval} are widely adopted and proven effective. ICL enables LLMs to infer task-specific patterns through representative examples embedded in prompts, allowing the model to mimic the desired output without additional training. RAG, on the other hand, augments the LLM's domain knowledge by retrieving relevant references from curated external databases and integrating them into the input prompts. Both techniques allow LLMs to adapt to new domains without extensive retraining efforts, making them highly suitable for developing LLM-based design script generation tools. However, as summarized in \tablename~\ref{tab:existing_works}, existing works focus exclusively on traditional electronic circuits, leaving the automation of PIC design scripting still unexplored.

\subsection{LLM-based PIC Design Scripting Challenges}
\label{sec:challenge}
The PIC design flow exhibits unique characteristics and introduces additional complexity, which significantly diminishes the effectiveness of existing solutions in the EDA domain. As an emerging field, it lacks large-scale, high-quality datasets, rendering training-based methods impractical. However, unlike the TCL scripts commonly used in conventional EDA flows \cite{ousterhout1993introduction}, PIC design scripts are typically written in Python, a language in which LLMs have demonstrated strong proficiency \cite{zan2025multi}. 
This makes training-free methods, particularly RAG, well-suited for PIC design scripting, as it enables LLM to effectively combine inherent Python programming capabilities with design tool knowledge retrieved from external databases, thereby generating executable and functionally correct PIC design scripts. 

Nevertheless, existing RAG-based tools are tailored for traditional circuit design flows \cite{liu2023chipnemo,liu2025layoutcopilot} and basically adopt the general RAG pipeline without sufficient optimization for specific task scenarios. As a result, these solutions exhibit limited transferability to LLM‑based PIC design script generation. As described in Section~\ref{sec:flow}, a typical PIC design flow requires coordinating multiple design tools from different vendors with heterogeneous APIs. 
This necessitates a script generation framework capable of handling complex scripting tasks involving multiple functional steps, while also integrating a tailored RAG pipeline to precisely extract tool‑specific knowledge. Moreover, the rapid evolution of the PIC ecosystem requires that it possesses strong scalability to accommodate the continuous emergence and iteration of diverse design tools. Therefore, it is imperative to optimize the RAG pipeline to better meet the inherent precision and adaptability requirements of retrieval in this domain, and building upon this foundation, specifically design an LLM‑based script generation framework to assist PIC design.

\subsection{Task Description}
In this work, we focus on leveraging LLMs to \textbf{assist PIC design by automatically generating design scripts} from natural language descriptions, rather than using LLMs to \textit{directly design PICs}. We formalize the PIC design script generation task as follows:
\begin{itemize}[nosep, leftmargin=*]
\item Given a natural language description of a PIC design scripting task, the goal is to generate an executable and functionally correct script that fully satisfies the task requirements.
\end{itemize}

%% file: Framework.tex
\section{PICopilot Framework}

\subsection{Framework Overview}
Figure~\ref{fig:pic_copilot} presents an overview of PICopilot, an automated framework for PIC design script generation. We adopt a multi-agent architecture with a feedback mechanism for scalability and robustness, which enables the seamless integration of new agents as PIC design tools rapidly evolve. In step \ding{202}, the PIC designer provides a natural language description of a script-writing task. The Task Planner Agent analyzes the task instruction, decomposes it into subtasks of different functional domains if the task is composite, and routes them to the corresponding Function-specific Script Generator Agents (\ding{203}). Simultaneously, the planning information is transmitted to the Script Synthesizer Agent to guide subsequent code synthesis. In step \ding{204}, these generators produce scripts with different specific functions leveraging our tailored RAG pipeline and forward them to the Script Synthesizer Agent. The synthesizer integrates all individual scripts into a unified final version, which is then delivered to the Script Evaluator Agent and displayed to the designer (\ding{205}). In step \ding{206}, the PIC designer can provide modification instructions, which are also forwarded to the evaluator. The Script Evaluator Agent jointly analyzes the generated script and any designer feedback to determine whether revisions are needed. If so, an adaptive feedback loop is triggered (\ding{207}) to iteratively refine the script. Finally, PICopilot outputs the finalized script in step \ding{208}.

\subsection{General Agent Design Techniques}

\begin{figure}[t]
% \vspace{-8pt}
% \vspace*{0.65\baselineskip}
\centerline{\includegraphics[width=\linewidth]{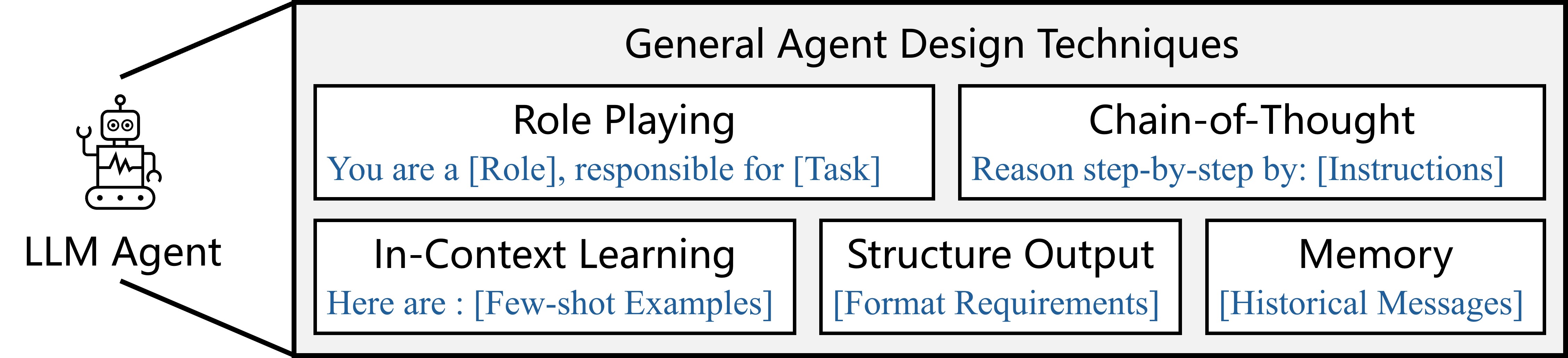}}
\vspace*{-0.65\baselineskip}
\caption{An illustration of our agent design techniques.}
\label{fig:agent_strategy}
% \vspace*{0.2\baselineskip}
% \vspace*{-0.2\baselineskip}
\end{figure}

As illustrated in Figure~\ref{fig:agent_strategy}, each LLM agent in PICopilot adopts a suite of general techniques to enhance performance in addition to its equipped LLM, offering advantages over direct LLM invocation. The role-playing technique is employed to explicitly define each agent's functional role and task scope, enhancing coordination and consistency throughout the PIC design script generation process. To enhance task adaptability, we provide task-specific few-shot examples for each agent to enable effective ICL. Chain-of-thought (CoT) prompting \cite{wei2022chain} is further utilized to strengthen the reasoning ability of the LLM agents. This technique guides them to complete tasks through step-by-step logical deduction, improving both success rates and interpretability. Structured outputs in JSON format are enforced to ensure that agents communicate in standardized, machine‑readable formats, which reduces parsing ambiguity and integration errors. Furthermore, each agent maintains a memory of previous messages, which preserves contextual continuity and facilitates coherent revision during feedback loop iterations.

% \begin{figure}[htbp]
% % \vspace{-8pt}
% \vspace*{-0.65\baselineskip}
% \centerline{\includegraphics[width=\linewidth]{figure/agent_strategy_new.jpg}}
% % \vspace{-9pt}
% \vspace*{-0.65\baselineskip}
% \caption{An illustration of our agent design techniques.}
% \label{fig:agent_strategy}
% % \vspace{-16pt}
% \vspace*{-0.65\baselineskip}
% \end{figure}

\subsection{Task Planner Agent}
% long
The Task Planner Agent serves as the central coordinator of PICopilot, which is responsible for interpreting and organizing scripting tasks from the PIC designer by invoking an LLM with the aforementioned optimization techniques. Upon receiving a task description, it first analyzes the design intent and determines whether the task is composite or function-specific. For composite tasks spanning multiple functional domains, the agent decomposes them into a set of subtasks, each corresponding to a script generator for a specific function (e.g., layout design or design rule check (DRC)). It then performs task description rewriting to eliminate potential ambiguities and supplement missing details, and routes each rewritten task to the assigned script generator. Concurrently, it transmits task planning details, including task type and subtask information, to the synthesizer to guide final script generation. 
By orchestrating the entire workflow and decomposing complex tasks across specialized functional domains, the Task Planner Agent enhances the processing performance and scalability of the entire framework.

% \begin{figure}[htbp]
% \vspace*{0.55\baselineskip}
% \centerline{\includegraphics[width=\linewidth]{figure/retrieval_Idea_new_new_resize.jpg}}
% \vspace*{-0.65\baselineskip}
% \caption{An illustration of the retrieval paradigm of the PIC designer and PICopilot.}
% \label{fig:retrieval}
% \vspace*{0.75\baselineskip}
% \end{figure}

\subsection{Function-specific Script Generator Agents}
PICopilot implements PIC design script generation through a set of Function-specific Script Generator Agents rather than a monolithic generator, which ensures high scalability, reconfigurability, and excellent scripting capabilities. As shown in Figure~\ref{fig:pic_copilot}, each generator utilizes an agentic architecture with our specifically designed RAG pipeline, where multiple sub-agents collaborate to generate correct PIC design scripts by combining the LLM's inherent programming skills with scripting knowledge retrieved from external databases.

\begin{figure}[t]
% \vspace*{-0.35\baselineskip}
\centerline{\includegraphics[width=\linewidth]{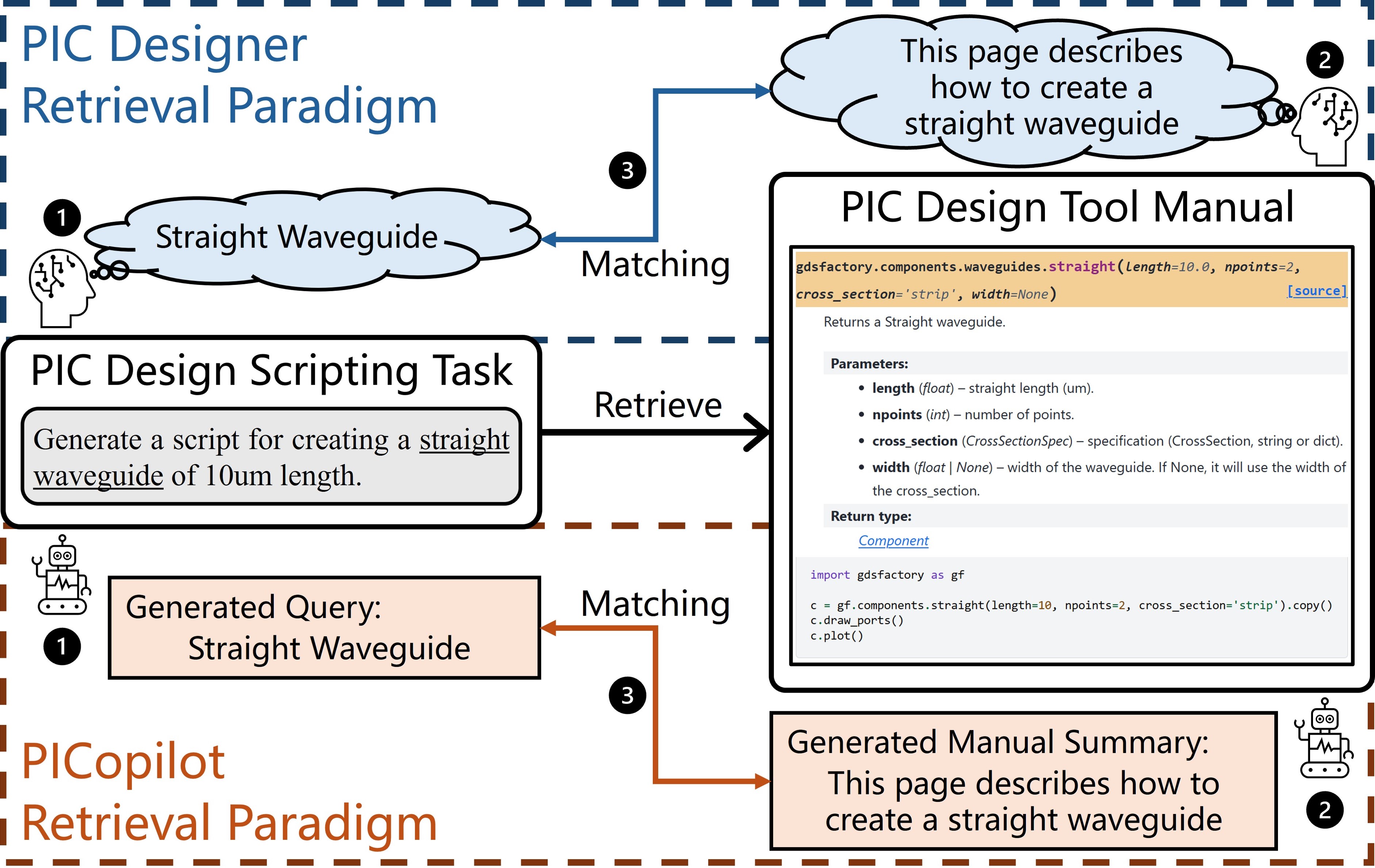}}
\vspace*{-0.65\baselineskip}
\caption{An illustration of the retrieval paradigm of the PIC designer and PICopilot.}
\label{fig:retrieval}
% \vspace*{-0.65\baselineskip}
\end{figure}

% V5
To address the challenges of LLM-based PIC design scripting discussed in Section~\ref{sec:challenge}, PICopilot adopts a tailored RAG pipeline instead of the general one, achieving superior performance by aligning with the practical retrieval paradigm of PIC designers. As depicted in Figure~\ref{fig:retrieval}, when writing PIC design scripts, designers typically first formulate search queries in their minds based on the task (\ding{202}). Subsequently, they consult the design tool API manuals, mentally summarize the technical content (\ding{203}), and match these summaries with their queries to identify relevant references (\ding{204}). This paradigm achieves high retrieval precision by bridging the semantic gap between different text modalities (e.g., natural language and code) and filtering out redundant information contained in the original documents. Inspired by this process, PICopilot's script generators adopt a similar retrieval paradigm. As illustrated in Figure~\ref{fig:retrieval}, this paradigm performs matching between queries and summaries generated by the LLM that emulates human PIC designers, rather than directly matching original task descriptions with tool manual pages used in existing methods, thereby taking advantage of the human paradigm to effectively enhance retrieval performance.

\begin{figure*}[htbp]
% \vspace*{-0.65\baselineskip}
\centerline{\includegraphics[width=0.85\linewidth]{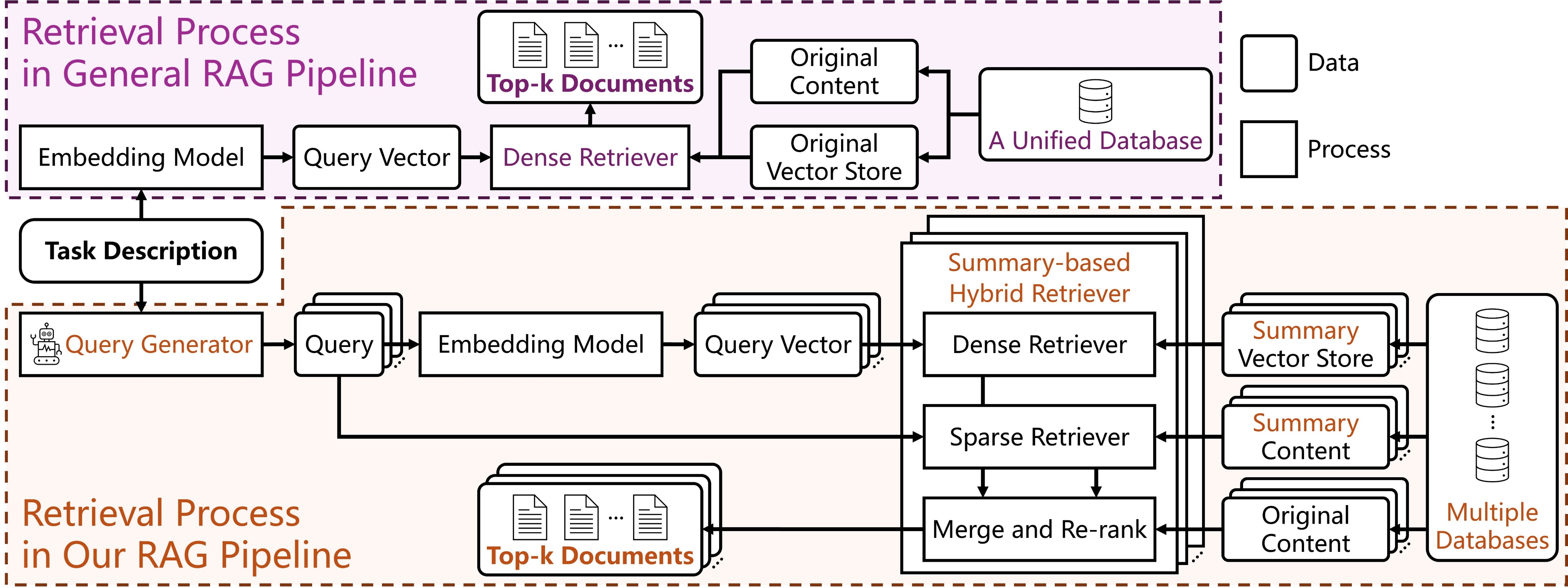}}
\vspace*{-0.67\baselineskip}
\caption{An illustration of the retrieval process in the general RAG pipeline and our RAG pipeline.}
\label{fig:retrieval_p}
\vspace*{-0.65\baselineskip}
\end{figure*}
% V5
Building on this paradigm, as illustrated in Figure~\ref{fig:retrieval_p}, we implement a customized RAG pipeline that integrates a Query Generator Agent and multiple summary-based hybrid retrievers to perform the retrieval process. The query generator emulates the query‑formulation behavior of human designers, while each retriever extracts relevant reference content from its corresponding tool manual database, adhering to the query-summary retrieval paradigm. Then, the Programmer Agent leverages the retrieved information in conjunction with the LLM's inherent programming proficiency to generate function-specific scripts. In contrast, the general RAG pipeline typically employs a single dense retriever that performs retrieval by calculating embedding similarities (Figure~\ref{fig:retrieval_p}). This conventional approach lacks optimization for our application scenario and suffers from poor cross-modal matching and interference caused by redundant information, rendering it unsuitable for direct application in the LLM-based PIC design scripting task.

\begin{figure}[t]
\vspace*{0.55\baselineskip}
\centerline{\includegraphics[width=0.82\linewidth]{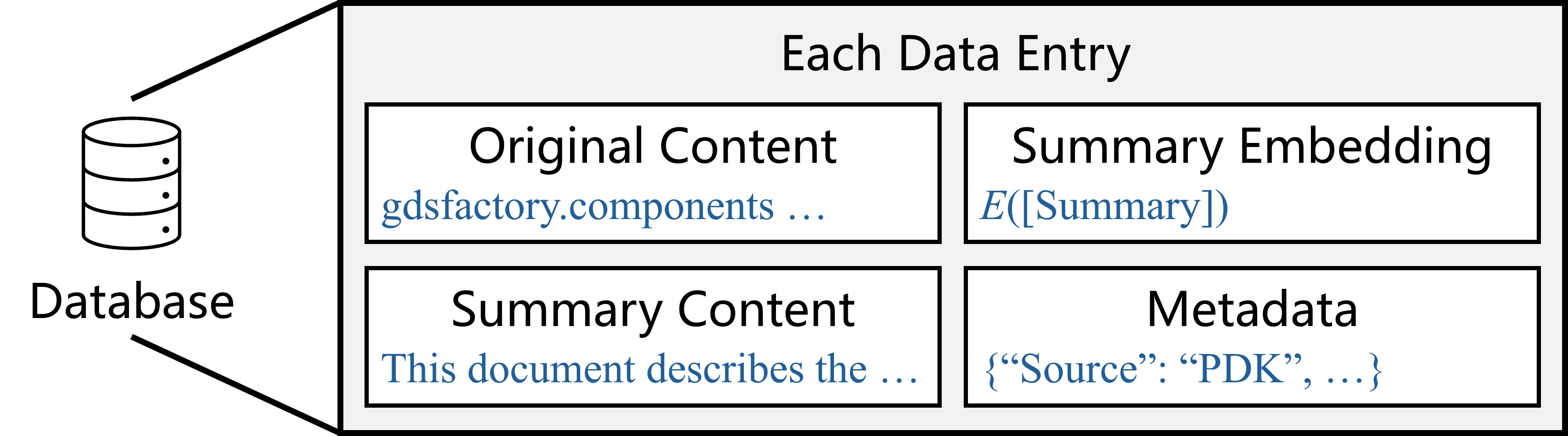}}
\vspace*{-0.65\baselineskip}
\caption{An illustration of our database design.}
\label{fig:database}
% \vspace{-16pt}
% \vspace*{0.1\baselineskip}
\vspace*{-0.2\baselineskip}
\end{figure}

\subsubsection{Query Generator Agent}

The Query Generator Agent emulates the human PIC designer’s query formulation, transforming task descriptions into multiple targeted retrieval queries. After receiving the task input processed by the Task Planner Agent, it generates a set of concise retrieval queries aligned with the design intent by invoking an LLM with the prompt that includes detailed instructions and few-shot examples. Since each script generator can contain multiple databases and retrievers for enhanced accuracy and scalability, the agent also selects the target database for each query and routes it to the corresponding retriever. By effectively mitigating ambiguity and improving query quality, this agent significantly enhances the performance of our tailored RAG pipeline.

\subsubsection{Tool Manual Databases} \label{sec:database}

To ensure accurate and flexible retrieval, each Function-specific Script Generator in PICopilot employs a multi-database design, as shown in Figure~\ref{fig:retrieval_p}, rather than maintaining a single unified database used in previous methods. In practice, each function-specific step in the PIC design flow typically necessitates access to different tool knowledge bases. For instance, scripting tasks for layout design simultaneously rely on both the layout tool manual and the process design kit (PDK) tool manual. Consolidating all knowledge from diverse sources into a single database often introduces semantic interference, such as API naming conflicts and conceptual ambiguities, which ultimately compromise retrieval precision. Furthermore, a unified database is difficult to maintain and scale, since updating existing documents or adding new documents may require costly re-indexing and re-embedding operations. To overcome these limitations, PICopilot adopts a multi-database structure within each generator, improving retrieval performance and enhancing scalability.

% \begin{figure}[htbp]
% \vspace*{-0.45\baselineskip}
% \centerline{\includegraphics[width=0.82\linewidth]{figure/database_new.jpg}}
% \vspace*{-0.65\baselineskip}
% \caption{An illustration of our database design.}
% \label{fig:database}
% % \vspace{-16pt}
% \vspace*{-0.45\baselineskip}
% \end{figure}

Each database in our framework is constructed from a specific tool manual, and we propose a general construction workflow comprising data cleaning, structured segmentation, and summary generation. During the cleaning step, we remove non-textual elements (e.g., images) and retain only textual content. Manual content with hierarchical chapter structures is divided into chunks according to the smallest units to maintain coherence and semantic integrity. For API documentation, content is segmented at the granularity of individual APIs. Oversized chunks are further subdivided and annotated with metadata to preserve contextual traceability. Each chunk is then summarized using an LLM to generate concise content summaries that unify semantics and eliminate redundancy, just like a human designer. Additionally, we incorporate a human expert verification step to ensure the factual accuracy of summaries. Although this step is time-consuming, it is a one-time investment and crucial for avoiding issues caused by LLM hallucinations and randomness. As shown in Figure~\ref{fig:database}, the final database entries integrate the original content, summary content, summary embedding, and metadata, which facilitates the subsequent high-precision retrieval.

\subsubsection{Summary-based Hybrid Retriever}

Given that each Function-specific Script Generator Agent employs a multi-database structure, we deploy a set of summary-based hybrid retrievers, each dedicated to a specific database. This distributed architecture offers superior scalability, as new databases and retrievers can be seamlessly integrated without affecting existing components. 
As illustrated in Figure~\ref{fig:retrieval_p}, for each query $q$ generated by the Query Generator Agent, the designated retriever processes the summary set $S=\{s_1, s_2,...,s_n\}$ in its associated database rather than the original content. By utilizing our tailored retrieval algorithm, the retriever identifies relevant summaries and retrieves the corresponding original text chunks as references for subsequent generation of PIC design scripts.

Algorithm~\ref{alg:retrieval} details our designed retrieval process, and we employ a hybrid strategy because both semantic matching and keyword matching are essential for high‑precision retrieval in PIC design script generation. Semantic similarity enables retrievers to effectively identify relevant content based on query intent and meaning, while keyword precision ensures accurate and efficient retrieval of elements that are highly dependent on lexical form, such as API names. Therefore, as shown in Figure~\ref{fig:retrieval_p}, each of our retrievers combines both a dense retriever and a sparse retriever, fully leveraging the advantages of both matching modes. The dense retriever first encodes the query $q$ into the same embedding space as the precomputed summary embeddings $\{E(s_i)\}$ and calculates semantic similarity scores using the cosine similarity formula:
\begin{equation}
\vspace*{-0.15\baselineskip}
% {sim}_{\mathrm{dense}}(q, s_i) = 
% {sim}(q, s_i) = 
{score}_{\mathrm{dense}}(q, s_i) = 
\frac{E(q) \cdot E(s_i)}{\|E(q)\| \, \|E(s_i)\|}
\label{eq:cosine}
% \vspace*{-0.05\baselineskip}
\end{equation}
where $E(q)$ and $E(s_i)$ denote the embeddings of the query and summary. Summaries are then ranked in descending order of the score, and the top-$k$ ones are selected as $S^{\mathrm{dense}}_k$. The sparse retriever computes lexical relevance between the query and each summary using the classical BM25 formula as follows:
\begin{equation}
% \vspace*{-0.1\baselineskip}
% \mathrm{score}_{\mathrm{BM25}}(q, s_i) =
{score}_{\mathrm{sparse}}(q, s_i) =
\sum_{t \in q} \mathrm{IDF}(t) \cdot
% \frac{f(t, s_i) \, (k_1 + 1)}
% {f(t, s_i) + k_1 \left(1 - b + b \frac{|s_i|}{\mathrm{avgsl}}\right)},
\frac{TF(t, s_i) \, (k_1 + 1)}
{TF(t, s_i) + k_1 \left(1 - b + b \frac{|s_i|}{\mathrm{avgsl}}\right)}
\label{eq:bm25}
\vspace*{-0.1\baselineskip}
\end{equation}
% Long
where $TF(t,s_i)$ represents the frequency of term $t$ in summary $s_i$ and $IDF(t)$ is the inverse document frequency of $t$. $|s_i|$ is the summary length, and $\mathrm{avgsl}$ is the mean summary length. $k_1$ and $b$ are empirical parameters, and we retain their default values of 1.5 and 0.75. The sparse retriever then ranks summaries by ${score}_{\mathrm{sparse}}$ and extracts the top-$k$ results, denoted as $S^{\mathrm{sparse}}_k$. After both retrievers produce their ranked lists, our retriever merges $S^{\mathrm{dense}}_k$ and $S^{\mathrm{sparse}}_k$, removes duplicates, and re-scores all candidate summaries by using the weighted reciprocal ranking fusion (RRF) strategy as follows:
\begin{equation}
% \vspace*{-0.1\baselineskip}
% \mathrm{score}_{\mathrm{RRF}}(q, s_i) =
\mathrm{score}(q, s_i) =
\sum_{r \in \{\mathrm{dense},~\mathrm{sparse}\}}
\frac{w_r}{c + \mathrm{rank}_r(s_i)}
\label{eq:rrf}
\vspace*{-0.1\baselineskip}
\end{equation}
% short
where $\mathrm{rank}_r(s_i)$ is the rank of summary $s_i$ in the list generated by retriever $r$, and $c$ is a smoothing constant set to 60 by default. $w_r$ is the importance weight and we set $(w_{\mathrm{dense}}, w_{\mathrm{sparse}})=(0.7, 0.3)$ based on actual testing. This process integrates both semantic and lexical evidence into a unified score, ensuring a highly robust ranking. The top‑$k$ summaries are then selected to form the final retrieval set $S_k$, and their corresponding original text chunks are retrieved and passed to the Programmer Agent for script generation.

% \vspace*{-0.25\baselineskip}
\begin{algorithm}[t]
% \begin{spacing}{0.9}
\begin{spacing}{1}
\caption{Summary-based Hybrid Retrieval}
\label{alg:retrieval}
\begin{algorithmic}[1]
\State \textbf{Input:} query $q$, database summary set $S$, retrieved count $k$
\State \textbf{Output:} $k$ original document chunks
\State Construct $S^{\mathrm{dense}}_k$ with Equation~\eqref{eq:cosine}; \Comment{Dense retrieval}
\State Construct $S^{\mathrm{sparse}}_k$ with Equation~\eqref{eq:bm25}; \Comment{Sparse retrieval}
\State Merge $S^{\mathrm{dense}}_k$ and $S^{\mathrm{sparse}}_k$, and remove duplicates; \Comment{Merge}
\State Construct $S_k$ with Equation~\eqref{eq:rrf}; \Comment{Re-rank}
\State Retrieve the original document chunks corresponding to $S_k$;
\end{algorithmic}
\end{spacing}
\end{algorithm}
% \vspace*{-0.4\baselineskip}

% % \vspace*{-0.3\baselineskip}
% \begin{algorithm}[htbp]
% \begin{spacing}{0.9}
% % \begin{spacing}{1}
% \caption{Summary-based Hybrid Retrieval}
% \label{alg:retrieval}
% \begin{algorithmic}[1]
% \State \textbf{Input:} query $q$, database summary set $S$, threshold $k$
% \State \textbf{Output:} $k$ original document chunks
% \State Construct $S^{dense}_k$ with Equation~\eqref{eq:cosine}; \Comment{Dense retrieval}
% \State Construct $S^{sparse}_k$ with Equation~\eqref{eq:bm25}; \Comment{Sparse retrieval}
% % \State Construct $S^{dense}_k$ and $S^{sparse}_k$ with Equation~\eqref{eq:cosine} and Equation~\eqref{eq:bm25};
% \State Merge $S^{dense}_k$ and $S^{sparse}_k$, and remove duplicates; \Comment{Merge}
% \State Construct $S_k$ with Equation~\eqref{eq:rrf}; \Comment{Re-rank}
% \State Retrieve the original document chunks corresponding to $S_k$;
% % \State $G_{sorted} \gets$ Sort groups $g$ based on descending $H[g]$
% %     \For{each group $g_i \in G_{sorted}$}
% %         \State $B' \gets \text{ReduceBitwidth}(B, g_i)p$
% %         \If{$E(B') > 1.1 \cdot \epsilon_{\text{max}}$}
% %             \State $B' \gets \text{Compensate}(B', G_{sorted}[i+1:])$
% %         \EndIf
% %         \State $\text{UpdatePoolIfBetter}(P, B')$
% %     \EndFor
% % \EndFor
% \end{algorithmic}
% \end{spacing}
% \end{algorithm}
% \vspace*{-0.4\baselineskip}t

\subsubsection{Programmer Agent}
The Programmer Agent generates PIC design scripts in Python format based on the refined task description from the Task Planner Agent and the reference materials provided by the retrievers. By combining the powerful Python programming capabilities of LLMs with retrieved design tool scripting knowledge, it can generate scripts that correctly complete specified tasks.

\subsection{Script Synthesizer}
% Long
The Script Synthesizer Agent takes as input scripts generated by different Function-specific Script Generator Agents and task planning information provided by the Task Planner Agent. It synthesizes a coherent, executable final script from the inputs by invoking an LLM with a prompt, which incorporates step‑by‑step reasoning, detailed instructions, and tailored few‑shot examples. For non-composite tasks, it directly outputs scripts without LLM invocation.

\subsection{Script Evaluator}
The Script Evaluator Agent receives the script from the Script Synthesizer Agent, supplemented with optional feedback from the PIC designer. To facilitate accurate evaluation and feedback, the task description and each preceding agent's reasoning trace are also sent to it along the data flow. Although directly executing generated scripts for evaluation is common and effective, it is impractical because running PIC design tools is extremely time-consuming (e.g., a typical electromagnetic simulation of a single device can take several hours). Therefore, we design this agent to perform evaluation via an LLM equipped with a static code checker and a custom check library, leveraging the model's strong capabilities in comprehension, reasoning, and programming.

The evaluator assesses the script in terms of its correctness as well as alignment with the intended task, and determines whether revisions are needed. Specifically, it first invokes the static code checker, which is implemented in Python with the AST \cite{AST} and Pyflakes \cite{Pyflakes} libraries, to check the script for syntax and logic issues without executing it. The diagnostic messages returned by the checker are systematically organized into the LLM prompt, providing the necessary information for evaluation. The agent also integrates all the check suggestions from the custom check library into the final prompt, which guides the LLM to focus on error-prone parts of the generated script during evaluation. We establish this library by collecting common errors found in generated PIC design scripts and rewriting them as check prompts. Its content can also be customized by users. Ultimately, the LLM is invoked to evaluate the script and provide feedback through our crafted prompt, which comprises the agent’s input, information from the checker and library, detailed instructions and few-shot cases.

As depicted in Figure~\ref{fig:pic_copilot}, if modifications are required, the agent adaptively generates targeted revision prompts and sends them to the corresponding agents. The script generation process then restarts at the first agent receiving feedback, and each agent updates its output based on the new input and historical messages stored in its memory. The loop continues until the evaluator determines that no revision is needed or the maximum iteration count $N_f$ is reached. This adaptive feedback mechanism can improve the success rate of PICopilot in generating scripts and mitigate reliability issues caused by the LLM's inherent hallucinations and randomness.

%% file: Experiment.tex
\begin{table*}[t]
\centering
% \small
\footnotesize
% \scriptsize
\setlength{\tabcolsep}{5pt}
\renewcommand{\arraystretch}{1.1}
\caption{Comparison of the PIC design script generation results between PICopilot and baseline methods.}
% \vspace*{-0.65\baselineskip}
\vspace*{-0.85\baselineskip}
\label{tab:main_result}
\begin{tabular}{cc|c|c|c|c|c|c|c}
\toprule
\multicolumn{2}{c|}{\multirow[c]{2}{*}{\textbf{Task Set}}} & \multicolumn{2}{c|}{\textbf{Qwen3-Coder}} & \multicolumn{2}{c|}{\textbf{DeepSeek-V3.2}} & \multicolumn{2}{c|}{\textbf{GPT-5}} & \multirow[c]{2}{*}{\textbf{PICopilot}} \\
% & & \textbf{Zero-shot} & \textbf{ICL \& RAG} & \textbf{Zero-shot} & \textbf{ICL \& RAG} & \textbf{Zero-shot} & \textbf{ICL \& RAG} & \\
& & Zero-shot & ICL \& RAG & Zero-shot & ICL \& RAG & Zero-shot & ICL \& RAG &  \\
% & & \textbf{Zero-shot}$^\dagger$ & \textbf{ICL \& RAG}$^\dagger$ & \textbf{Zero-shot} & \textbf{ICL \& RAG} & \textbf{Zero-shot} & \textbf{ICL \& RAG} & \\
\midrule
\multirow[c]{3}{*}{\textbf{Basic}} & Pass@1 & 35.6 & 53.9 & 31.7 & 50.0 & 26.7 & 51.7 & \textbf{98.3} \\
& Pass@5 & 40.4 & 58.2 & 33.3 & 50.0 & 43.7 & 56.1 & \textbf{100.0} \\
& \#Solved & 5 & 7 & 4 & 6 & 7 & 7 & \textbf{12} \\
\midrule
\multirow[c]{3}{*}{\textbf{Medium}} & Pass@1 & 10.0 & 32.2 & 11.1 & 47.8 & 4.4 & 42.8 & \textbf{96.1} \\
 & Pass@5 & 14.5 & 33.3 & 16.0 & 58.3 & 13.5 & 62.7 & \textbf{100.0} \\
& \#Solved & 2 & 4 & 2 & 7 & 3 & 8 & \textbf{12} \\
\midrule
\multirow[c]{3}{*}{\textbf{Advanced}} & Pass@1 & 7.5 & 20.3 & 4.7 & 36.1 & 3.1 & 25.8 & \textbf{90.6} \\
& Pass@5 & 8.3 & 25.0 & 6.6 & 42.8 & 6.5 & 42.8 & \textbf{99.8} \\
& \#Solved & 2 & 6 & 2 & 11 & 2 & 12 & \textbf{24} \\
\midrule
\multirow[c]{3}{*}{\textbf{Total}} & Pass@1 & 15.1 & 31.7 & 13.1 & 42.5 & 9.3 & 36.5 & \textbf{93.9} \\
& Pass@5 & 17.9 & 35.4 & 15.6 & 48.5 & 17.6 & 51.1 & \textbf{99.9} \\
& \#Solved & 9 & 17 & 8 & 24 & 12 & 27 & \textbf{48} \\
\bottomrule
% \multicolumn{9}{l}{\textbf{Zero-shot}: generate scripts directly from task descriptions; \textbf{ICL \& RAG}: generate scripts with ICL and the general RAG pipeline.}
\multicolumn{9}{l}{Zero-shot: generate directly from task descriptions; ICL \& RAG: generate with ICL and the general RAG pipeline.}
% \multicolumn{9}{l}{$\dagger$ \textbf{Zero-shot}: generate scripts directly from task descriptions; \textbf{ICL \& RAG}: generate scripts with ICL and the general RAG pipeline.}
\end{tabular}
% \vspace*{-0.65\baselineskip}
\vspace*{-1.2\baselineskip}
\end{table*}

\section{Experimental Results}
\subsection{Experimental Setup}

\subsubsection{Implementation}
We implement PICopilot in Python with the LangChain framework \cite{langchain}. The Programmer Agent is powered by Qwen3-Coder due to its advanced programming ability \cite{qwen3technicalreport}, while other agents use the general-purpose model DeepSeek-V3.2 \cite{deepseek2025deepseek}. For the RAG pipeline, we adopt EmbeddingGemma \cite{vera2025embeddinggemma}, an open-source embedding model recognized for its high performance in Python-related retrieval tasks \cite{li2407coir}. Following the method in Section~\ref{sec:database}, we establish multiple tool manual databases based on commonly used PIC design tools covering various functions. These database summaries are generated via DeepSeek-V3.2 and verified by a PIC design expert to ensure accuracy. Notably, our database construction method is generalized, enabling users to build databases based on any PIC design tool manual. We set the top-$k$ retrieved documents per query to $k=5$, and the maximum iteration count $N_f=3$. To maintain fairness and eliminate human bias, no designer feedback is provided in any experiment. All experiments are conducted on a Linux machine with an Intel i7-13700 CPU and 128GB RAM, and all LLMs are invoked via APIs.

\begin{table}[t]
% \vspace*{-0.65\baselineskip}
% \vspace*{0.4\baselineskip}
\centering
\caption{PIC design script benchmark information.}
\vspace*{-0.65\baselineskip}
% \small
\footnotesize
\renewcommand{\arraystretch}{1.1}
\setlength{\tabcolsep}{4.6pt}
\begin{tabular}{c|c|c}
\toprule
\textbf{Task Set} & \textbf{Num.} & \textbf{Description}\\
\midrule
Basic & 12 & \makecell[l]{\textbf{single} domain; \textbf{$<$10} lines of core code. \\ \textit{(e.g., create a layout of ... by ... (Layout Design).)}} \\
\midrule
Medium  & 12 & \makecell[l]{\textbf{single} domain; \textbf{10--50} lines of core code. \\ \textit{(e.g., perform custom FDTD simulation on ... by ...,} \\ \textit{process results by ..., and export to ... (Simulation).)}} \\
\midrule
Advanced  & 24 & \makecell[l]{\textbf{multiple} domains; \textbf{$>$50} lines of core code. \\ \textit{(e.g.,  create a layout of ..., perform custom DRC by} \\ \textit{..., and extract ... to ... via default FDTD simulation} \\ \textit{(Layout Design + DRC + Simulation).)}} \\
\bottomrule
\end{tabular}
\label{tab:benchmark}
% \vspace*{-0.65\baselineskip}
\end{table}

\subsubsection{Baseline Methods}
To comprehensively evaluate the effectiveness of PICopilot, we select three representative and commonly used LLMs as baselines: GPT-5 \cite{GPT5}, DeepSeek-V3.2, and Qwen3-Coder. The first two are state‑of‑the‑art (SOTA) commercial and open‑source general-purpose models, while Qwen3-Coder is a leading model specialized in programming. Each LLM is evaluated under two distinct settings: (1) zero-shot generation: generate scripts directly from task descriptions, and (2) enhanced generation with ICL and RAG: generate scripts using our tailored prompt template with ICL and the general RAG pipeline in existing methods featuring a dense retriever and a unified database. To ensure fair comparison, their retrieved document number $k$ is set to match the total number of documents retrieved by our RAG pipeline for each task.

\subsubsection{Benchmark}
Due to the absence of publicly available benchmarks, we establish a comprehensive one summarized in Table~\ref{tab:benchmark}. It consists of 48 commonly used script generation tasks carefully selected from actual PIC design flows. Representative examples are listed in the table with some custom task-specific descriptions omitted due to space limitations. Each task has a ground-truth script that is written by a PIC designer and verified by running it and obtaining its output.
We divide these tasks into three difficulty levels based on the functional domains involved (e.g., layout design and simulation) and the amount of core code (excluding task-irrelevant code like library imports) required in the ground truth. To prevent fairness issues caused by test task leakage, extra scripting tasks are utilized as few-shot cases in our LLM prompts.

\subsubsection{Metrics}
% short
We adopt ‘Pass@k’ (k=1, 5) \cite{chen2021evaluating} as our main evaluation metric, which has been widely adopted in evaluating code generation tasks. It represents the probability that at least one of $k$ independent code generations is correct, with a higher value indicating a higher success rate and better performance. For PICopilot and all baselines, we perform $n=15$ independent generation trials per test task and calculate it by $\mathrm{Pass@k}=1 - {\binom{n-c}{k}}/{\binom{n}{k}}$,
where $c$ denotes the number of successful trials. The success of a trial is determined by executing the generated script. If the script's execution result is functionally identical to the test task's ground truth, the trial is considered successful; otherwise, it is considered a failure.

Additionally, we introduce ‘\#Solved’ metric to quantify overall task completion status, defined as the total number of ‘solved’ tasks. A task is considered ‘solved’ by the framework if it successfully generates correct scripts at least 3 times in 15 independent trials. This design ensures the metric reflects the framework's consistent script generation capability while excluding random successes.

\begin{table*}[h]
\centering
% \small
\footnotesize
\setlength{\tabcolsep}{5pt}
\renewcommand{\arraystretch}{1.1}
\caption{Overhead of PICopilot and baseline methods.}
% \vspace*{-0.65\baselineskip}
\vspace*{-0.85\baselineskip}
\label{tab:overhead}
\begin{tabular}{cc|c|c|c|c|c|c|c|c}
\toprule
\multicolumn{2}{c|}{\multirow[c]{2}{*}{\textbf{}}} & \multicolumn{2}{c|}{\textbf{Qwen3-Coder}} & \multicolumn{2}{c|}{\textbf{DeepSeek-V3.2}} & \multicolumn{2}{c|}{\textbf{GPT-5}} & \multirow[c]{2}{*}{\textbf{PICopilot}} & \multirow[c]{2}{*}{\makecell[c]{\textbf{Designer} \\ \textbf{Reference}}} \\

& & Zero-shot & ICL \& RAG & Zero-shot & ICL \& RAG & Zero-shot & ICL \& RAG & & \\

\midrule
\multirow[c]{2}{*}{\textbf{LLM}} & API Calls & 1 / 1 & 1 / 1 & 1 / 1 & 1 / 1 & 1 / 1 & 1 / 1 & 6 / 12 & - \\
 & Cost ($\times10^{-2}$ \$) & 0.11 / 0.40 & 0.26 / 0.57 & 0.08 / 0.23 & 0.27 / 0.44 & 1.38 / 2.60 & 1.41 / 2.50 & 0.67 / 1.42 & - \\
 \midrule
 \multirow[c]{2}{*}{\textbf{Time}} & LLM (s)& 10.69 / 32.89 & 9.56 / 34.16 & 51.45 / 142.51 & 41.96 / 123.72 & 24.69 / 53.48 & 12.94 / 35.41 & 35.88 / 113.22 & - \\
 & Total (s) & 10.69 / 32.89 & 9.70 / 34.35 & 51.45 / 142.51 & 42.10 / 123.91 & 24.69 / 53.48 & 13.07 / 35.57 & 36.02 / 113.46 & 1200 / 3600 \\
\bottomrule
\multicolumn{10}{l}{API Calls: average/maximum number of LLM API calls per task; Cost ($\times10^{-2}$ \$): average/maximum cost of LLM API per task in US dollars.} \\
\multicolumn{10}{l}{LLM (s): average/maximum LLM call time per task in seconds; Total (s): average/maximum total time per task in seconds.}
\end{tabular}
% \vspace*{-0.65\baselineskip}
\vspace*{-1.2\baselineskip}
\end{table*}

\subsection{Main Results}
We evaluate PICopilot against the baseline methods, with results listed in \tablename~\ref{tab:main_result}. The results show that PICopilot consistently outperforms all baseline methods across all difficulty levels of the PIC script generation tasks. It successfully solves all 48 tasks and achieves the highest scores in both ‘Pass@1’ and ‘Pass@5’ metrics, primarily due to our customized multi-agent design and RAG pipeline tailored for PIC design script generation.

% short
In contrast, even the SOTA programming-specific and general-purpose LLMs exhibit poor performance when directly generating scripts from task descriptions. As PIC is an emerging field, relevant data is lacking in LLM training corpora. This results in current LLMs having little expertise in writing PIC design scripts, leading to their poor performance on our tasks. Although ICL and RAG can supplement LLMs with related knowledge, their improvements remain limited mainly due to suboptimal retrieval performance caused by the simple retrieval paradigm and single-database structure. Furthermore, their single-agent setups lack the task adaptability and robustness inherent to PICopilot’s multi-agent architecture, preventing them from consistently generating correct scripts.

\vspace*{-0.2\baselineskip}
\subsection{Overhead Analysis}
% short
We also evaluate the overhead of each method in the above experiments, summarized in \tablename~\ref{tab:overhead}. Our analysis focuses on the LLM usage and time consumption, which are key metrics of the framework's practical feasibility, as high cost and latency are unacceptable. As shown in the table, PICopilot requires the most LLM API calls and incurs the third-highest cost when executing each script generation task, primarily due to our multi-agent design. However, this additional overhead is worthwhile since it significantly improves script generation performance (as shown in \tablename~\ref{tab:main_result}). In addition, the absolute cost of PICopilot remains negligible, averaging less than one cent per task, and is expected to decrease further as the LLM industry constantly evolves.

% v2
Regarding the temporal overhead of PIC design script generation, we list both the total latency and the specific time consumed by calling the LLM via API. For reference, we also provide the approximate average and maximum time for the designer to write ground-truth code during benchmark construction. Notably, this time is achieved through his proficiency in relevant APIs and Python, while designers lacking this expertise need to spend more time consulting manuals and programming. 
As demonstrated in \tablename~\ref{tab:overhead}, PICopilot generates scripts more efficiently than manual coding by the PIC designer. Compared to baseline methods that only invoke the LLM once, it also incurs no substantial latency penalties despite its complex architecture and increased LLM calls. 
Furthermore, we find that the vast majority of PICopilot's script generation latency stems from API-based LLM calls, which lie outside our optimization scope and are expected to be continuously improved with advancements in the LLM field. For methods using RAG, the time spent outside of LLM invocation is negligible. In summary, PICopilot maintains a highly acceptable overhead profile. It achieves superior performance without significantly increasing cost or latency compared to direct LLM calling, and its overhead is much lower than training-based methods that necessitate expensive high-performance servers.

\vspace*{-0.2\baselineskip}
\subsection{Ablation Study}
We further conduct an ablation study to validate the necessity and effectiveness of our proposed RAG pipeline and multi-agent architecture in PICopilot. To highlight their roles in handling complex PIC script generation tasks, we conduct experiments exclusively on the ‘Advanced’ task set, with results summarized in \tablename~\ref{tab:ablation}. ‘w/o Tailored RAG’ denotes replacing the tailored RAG pipeline in all script generators of PICopilot with the general one used in the baseline method. ‘w/o Multi-Agent’ means removing all agents except the script generator, leaving only a single unified generator with multiple databases to directly produce the final scripts. 

\begin{table}[t]
% \vspace*{-0.65\baselineskip}
% \vspace*{0.4\baselineskip}
\caption{Ablation experiments to analyze effects of our RAG pipeline and multi-agent architecture.}
\vspace*{-0.73\baselineskip}
\renewcommand{\arraystretch}{1.1}
\footnotesize
\setlength{\tabcolsep}{5pt}
\begin{center}
\begin{tabular}{l|c|c|c}
\toprule
\textbf{Method} & \textbf{Pass@1} & \textbf{Pass@5} & \textbf{\#Solved} \\ 
\midrule
PICopilot & \textbf{90.6} & \textbf{99.8} & \textbf{24} \\
PICopilot w/o Tailored RAG & 60.3 & 82.5 & 18 \\
PICopilot w/o Multi-Agent & 54.4 & 81.8 & 17 \\
\bottomrule
\end{tabular}
\label{tab:ablation}
\end{center}
\end{table}

% short
The results demonstrate that both the tailored RAG pipeline and the multi-agent architecture significantly enhance PICopilot's capability in PIC design script generation, and their removal results in noticeable performance degradation. Specifically, our tailored RAG pipeline achieves superior retrieval performance over the general one because it considers the PIC script writing characteristics and aligns with human designers' retrieval paradigm. This ensures that the Programmer Agent consistently obtains accurate reference materials during coding, substantially enhancing the validity and functional correctness of the generated scripts. Moreover, PICopilot's multi-agent architecture decomposes the original complex task into simple and explicit sub-tasks, reducing the complexity faced by each LLM and improving the success rate. Simultaneously, the feedback mechanism enabled by the multi-agent design effectively mitigates errors arising from LLM stochasticity and hallucinations. Consequently, these two designs are essential for PICopilot and significantly improve the success rate of PIC design script generation.

%% file: Conclusion.tex
\vspace*{-0.2\baselineskip}
\section{Conclusion}

In this paper, we present PICopilot, the first LLM-based framework designed for assisting PIC design via script generation. It pioneers the application of LLMs' powerful programming capabilities to automate labor-intensive and time-consuming script-writing tasks within PIC design flows, enabling human designers to focus on high-level innovation while significantly enhancing productivity. By employing a multi-agent architecture with a feedback mechanism and a tailored RAG pipeline, PICopilot achieves accurate and robust generation of
PIC design scripts. Experimental results demonstrate that our framework delivers superior script generation performance at a reasonable total cost and latency compared to existing LLM-based methods, contributing to the emerging PDA field.